\documentclass[12pt]{iopart}
\usepackage{iopams}
\usepackage{graphicx}
\usepackage{subfigure}
\usepackage{amssymb}

\begin{document}

\title{Effects of Dipole-Dipole Interaction on the Transmitted spectrum
of Two-level Atoms trapped in an optical cavity}

\author{Yuqing Zhang$^{1}$,  Lei Tan$^{1,2}$\footnote{E-mail : lei.tan@ucl.ac.uk}, Peter Barker$^{2}$}
\address{$^{1}$Institute of Theoretical Physics, Lanzhou
University, Lanzhou 730000, China \\ $^{2}$Department of Physics and Astronomy, University College London,
Gower Street, London WC1E 6BT, United Kingdom}

\begin{abstract}
The transmission spectrum of two dipole-dipole coupled atoms interacting with a
single-mode optical cavity in strong coupling regime is investigated theoretically for
the lower and higher excitation cases, respectively. The dressed states containing the
dipole-dipole interaction (DDI) are obtained by transforming the two-atom system into
an effective single-atom one. We found that the DDI can enhance the effects resulting
from the positive atom-cavity detunings but weaken them for the negative detunings cases
for lower excitation, which can promote the spectrum exhibiting two asymmetric peaks and
shift the heights and the positions of them. For the higher excitation cases, DDI can
augment the atomic saturation and lead to the deforming of the spectrum.
Furthermore, the large DDI can make the atom and the cavity decouple, making
a singlet of the normal-mode spectrum.
\end{abstract}


\section{Introduction}
Realization of the strong coupling regime (SCR) in cavity quantum
electrodynamics has opened a new vision for the  exploration of
quantum mechanics~\cite{Berman,Miller}. SCR is very attractive,
as photons emitted by atoms inside the cavity mode can be
reabsorbed and reemitted, etc., leading to Rabi oscillations~\cite{Thompson},
which gives rise to a normal-mode splitting in the eigenvalue
spectrum of the atom-cavity system~\cite{Leslie,Gea-Banacloche,Wu}.
The normal-mode splitting has been observed  with atomic beams passing a cavity
in  both the microwave regime~\cite{Agarwal,Brune,Bernardot} and
optical regime~\cite{Childs}, as well as with trapped atoms
in  optical cavities~\cite{Boca,Maunz,Klinner,Puppe}.
In experiment, the normal-mode splitting  is detected by probing
the transmitted spectrum in low excitation.
With increased  excitation, the  spectrum
presents hysteresis, and then forms a close structure.
When the atoms are saturated, they decouple from the cavity
and  only a single peak appears in the spectrum~\cite{Mielke,Gripp}.

However, few previous studies have considered the DDI,
even  through many works have been explored  atom-cavity systems
with  DDI~\cite{Goldstein,Nicolosi,Li,Gillet,Alharbi,Chen,Peng}.
In fact, the DDI can profoundly affect the light absorption and lead to the
shift of the atomic energy levels~\cite{Wang}.
The renormalization  of the  atomic resonance frequency due to the DDI
can result in the optical bistability of  the atomic system~\cite{Ben-Aryeh,Singh}.
So it is natural to expect that the transmitted spectrum may present some novel
characteristics due to  the DDI.
Fortunately, in recent years substantial progress towards the study about
solid materials~\cite{Scheibner,Hennessy} and
ultracold atoms~\cite{Phillips,Cirac,Schneble,Chu}
proves a good platform for the study of DDI.
In the this work, we go one step further and investigate
the transmitted spectrum of two dipole-dipole coupled atoms trapped and strongly
coupled to an optical resonator. The behavior  of spectrum in steady-state
is studied for a wide range of DDI intensity and atom-cavity detuning.
The relation and distinction of their effects on the spectrum both
in weak excitation limit and higher excitation  are also  explored.

The paper is organized as follows: Sec.2 presents the theoretical
model under consideration and provides the steady-state solution by
solving  the master equation. Sec.3 is devoted to the study of the
transmitted spectrum in the  weak excitation limit.
Sec.4 describes the structure characteristics of transmitted spectrum
for a strong  driving intensity. The effects of  both the detuning
and the DDI on the  spectrum are discussed.
Finally, we present our conclusions in Sec.5.

\section{Model}

We consider two identical dipole-dipole coupled  two-level atoms interacting with a
single-mode high-finesse optical cavity. The system
is pumped along the cavity axis by a coherent laser field of
frequency $\omega_{p}$ and an effective
amplitude $\eta$. The Hamiltonian for
the system in the rotating wave and electric dipole
approximations  is given by~\cite{Peng}
\begin{eqnarray}
H&=&-\Delta_{c}a^{\dagger}a-\sum_{k=1}^{2}[\Delta_{a}\sigma^{\dagger}_{k}\sigma_{k}-g(a^{\dagger}\sigma_{k}+a\sigma^{\dagger}_{k})]\nonumber\\
&+&J(\sigma_{1}\sigma_{2}^{\dagger}+\sigma_{2}^{\dagger}\sigma_{1})+\eta(a+a^{\dagger})\label{eq1}
\end{eqnarray}
where $\Delta_{c}=\omega_{p}-\omega_{c}$, $\Delta_{a}=\omega_{p}-\omega_{a}$.
$\omega_{c}$ and $\omega_{a}$ are the  resonance frequencies
of the atoms and the cavity field, respectively.
$a^{\dagger}$ and $a$ are the field  creation and annihilation
operators, $\sigma_{k}^{\dagger}$ and  $\sigma_{k}$ represent
the  raising and lowering operators of  the atom $k$ ($k=1,2$).
The first term of Hamiltonian (1) is the free Hamiltonian of cavity.
The atomic free Hamiltonian and the interaction Hamiltonian of
atoms and cavity with  coupling strength $g$ are shown in the second term.
The third term describes the DDI between atoms
and the last term is  the pump field  Hamiltonian.
The DDI   is defined in the form~\cite{Nicolosi}
\begin{eqnarray}
J=\frac{3}{4}(\Gamma_{0}c^{3}/\omega_{a}^{3}r^{3})(1-3\cos^{2}\varphi)
\end{eqnarray}
where $r$ is the distance  between the atoms and $\varphi$ is the atomic dipole moments
with respect to the interatomic axis. $\Gamma_{0}$ denotes the atomic spontaneous
emission rate in free space.
Here, we assume the dipole moments of the two atoms are parallel to each other and
are polarized in the direction perpendicular to the interatomic axis.
Then, $\cos\varphi=0$, the DDI intensity only depends  on the positions of the two atoms in the cavity.

Dissipations results from excitations spontaneous emission   and cavity photonic leakage can be taken
into account within the quantum master equation of  density matrix $\rho$.
It is expressed in the usual Lindblad form in  Born-Markov approximation($\hbar=1$)~\cite{Kastoryano}
\begin{eqnarray}
\dot{\rho}&=&-i[H,\rho]+L_{\kappa}\rho+L_{\gamma}\rho+L_{\gamma^{'}}\rho\nonumber\\
L_{\kappa}\rho&=&\kappa[2a\rho a^{\dagger}-a^{\dagger}a\rho-\rho a^{\dagger}a]\nonumber\\
L_{\gamma}\rho&=&\sum_{k=1}^{2}\gamma(2\sigma_{k}\rho \sigma^{\dagger}_{k}-
\sigma^{\dagger}_{k}\sigma_{k}\rho-\rho\sigma^{\dagger}_{k}\sigma_{k})\nonumber\\
L_{\gamma^{'}}\rho&=&\gamma^{'}(2\sigma_{1}\rho \sigma^{\dagger}_{2}-\sigma^{\dagger}_{1}\sigma_{2}\rho-\rho\sigma^{\dagger}_{1}\sigma_{2})\nonumber\\
&+&\gamma^{'}(2\sigma_{2}\rho \sigma^{\dagger}_{1}-\sigma^{\dagger}_{2}\sigma_{1}\rho-\rho\sigma^{\dagger}_{2}\sigma_{1})\label{eq2}
\end{eqnarray}

Here, the non-unitary parts $L_{\kappa}\rho$, $L_{\gamma}\rho$ and $L_{\gamma^{'}}\rho$
describe the coupling of the field mode and the atoms to the environment.
The coefficients $\kappa$ and $\gamma$ are the decay rates of the cavity field
and the atoms respectively.   The atom-atom cooperation
induced by their coupling with a common reservoir is given by $\gamma^{'}$~\cite{Nicolosi,Alharbi,Lehmberg},
which is important only when the atomic distances
are small  relative to the radiation wavelength.

The time evolution of the   operators expectation values for  the atom-cavity
system can be obtained with the master equation
\begin{equation}
\langle\dot{a}\rangle=i(\tilde{\Delta}_{c}\langle a\rangle-g\langle\sigma_{1}\rangle-g\langle\sigma_{2}\rangle-\eta)
\end{equation}
\begin{equation}
\langle\dot{\sigma}_{1}\rangle=i(\tilde{\Delta}_{a}\langle\sigma_{1}\rangle+g\langle a\sigma_{1z}\rangle+\tilde{J}\langle\sigma_{1}\rangle\langle\sigma_{2z}\rangle)\nonumber\\
\end{equation}
\begin{equation}
\langle\dot{\sigma}_{2}\rangle=i(\tilde{\Delta}_{a}\langle\sigma_{2}\rangle+g\langle a\sigma_{2z}\rangle+\tilde{J}\langle\sigma_{2}\rangle\langle\sigma_{1z}\rangle)\nonumber\\
\end{equation}
\begin{equation}
\langle\dot{\sigma}_{1z}\rangle=2ig(\langle a^{\dagger}\sigma_{1}\rangle-\langle a\sigma_{1}^{\dagger}\rangle)-2\gamma(1+\langle\sigma_{1z}\rangle)
\end{equation}
\begin{equation}
\langle\dot{\sigma}_{2z}\rangle=2ig(\langle a^{\dagger}\sigma_{2}\rangle-\langle a\sigma_{2}^{\dagger}\rangle)-2\gamma(1+\langle\sigma_{2z}\rangle)
\end{equation}
where $\tilde{\Delta}_{a}=\Delta_{a}+i\gamma$,  $\tilde{\Delta}_{c}=\Delta_{c}+i\kappa$,
and $\tilde{J}=J-i\gamma^{'}$.

Then we can get the steady state solutions of Eqs.(4)-(6) by setting
$\langle\dot{a}\rangle$$=$$\langle\dot{\sigma_{1}}\rangle$$=$$\langle\dot{\sigma_{2}}\rangle$
$=$$\langle\dot{\sigma}_{1z}\rangle$$=$$\langle\dot{\sigma}_{1z}\rangle$$=$$0$.

When $g$$\gg$$(\gamma,\kappa)$, the atom$-$cavity system reaches a strong coupling regime.
The new eigenstates of the system are  described by the  dressed states,
which are linear combination of pairs of  bare atom states and cavity field state.
However,  it is difficult to find out the dressed states of the atom-cavity system
with two dipole-dipole coupled atoms.
In fact, when the excitation of the atoms is very low, we can adopt the methods in~\cite{Nicolosi}
and simplify the two-atom system to an effective single atom system.
Then the effective form of Hamiltonian $H$ in Eq.(1) can be written as
\begin{eqnarray}
H_{eff}&=&-\Delta_{c}a^{\dagger}_{1}a-(\Delta_{a}-J)\sigma_{1}^{\dag}\sigma_{1}\nonumber\\
&+&\sqrt{2}g(a^{\dagger}\sigma_{1}+a\sigma_{1}^{\dag})+\eta(a+a^{\dagger}).
\end{eqnarray}
In the transformed Hamiltonian,  the dipole coupled atoms are denoted by two fictitious  atoms.
Only one of them couples to the field mode with  frequencies $\omega_{a}+J$ and
an effective coupling strength  $\sqrt{2}g$, but the other atom freely evolves  decoupling from the field.
As a result, the  dressed states of  the transformed system are similar to that of the single-atom system
\begin{eqnarray}
|0\rangle&=&|g\rangle|0\rangle,\nonumber\\
|n_{-}\rangle&=&\sin\frac{\theta_{n}}{2}|e, n-1\rangle-\cos\frac{\theta_{n}}{2}|g, n\rangle,\nonumber\\
|n_{+}\rangle&=&\cos\frac{\theta_{n}}{2}|e, n-1\rangle+\sin\frac{\theta_{n}}{2}|g, n\rangle.
\end{eqnarray}
where  $\sqrt{n}$ is a photon number state, $\theta_{n}=\arctan 2\sqrt{2}g\sqrt{n}/(\Delta+J)$,
$\Delta=\omega_{a}-\omega_{c}$ is the detuning between atom and field.
The  corresponding eigenenergies  are
\begin{eqnarray}
E_{0}&=&0,\nonumber\\
E_{n\pm}&=&\omega_{c}+\frac{\Delta+J}{2}\pm\frac{1}{2}\sqrt{(\Delta+J)^{2}+8g^{2}n}.
\end{eqnarray}

Spectrum of the first doublet of these states in a degenerate system
(for $\omega_{a}$$=$$\omega_{c}$) splits into two new resonances,
called normal-mode or vacuum-Rabi splitting.
Observation of the normal-mode splitting,
in fact, is also a benchmark signature
that a system has reached the SCR of cavity QED.

To investigate the steady state normal-mode spectrum,
we introduce the  intracavity photon number~\cite{Schuster},
\begin{equation}
\langle a^{\dagger}a\rangle_{0}=|\langle a\rangle_{0}|^{2}\label{eq2}
\end{equation}
which  is given by the modulus square of $\langle a\rangle_{0}$ and
is sufficient to calculate a spectrum of the coupled atoms-cavity system.

\section{Normal-mode spectrum in low excitation limit}

By choosing an appropriate  pump beam, we can keep a weak pump
intensity and thus a low atomic excitation  can be achieved.
In this condition, $\langle\sigma_{1z}\rangle$, $\langle\sigma_{2z}\rangle$$\rightarrow-1$,
so we can set  $\langle a\sigma_{1z}\rangle$$=$$\langle a\sigma_{2z}\rangle$$=$$-\langle a\rangle$.
Then the  steady-state solution of Eqs.(4),(5) and (6) and the steady state intracavity photon number can be given:

\begin{equation}
\langle a\rangle_{0}=\frac{\eta}{\tilde{\Delta}_{c}}\cdot\frac{1}{1-v}\label{eq2}
\end{equation}

\begin{equation}
\langle\sigma_{1}\rangle_{0}=\frac{\eta v}{g}\cdot\frac{1}{1-v}\label{eq2}
\end{equation}

\begin{equation}
\langle\sigma_{2}\rangle_{0}=\frac{\eta v}{g}\cdot\frac{1}{1-v}\label{eq2}
\end{equation}

\begin{equation}
\langle a^{\dagger}a\rangle_{0}=\frac{\eta^{2}}{|\tilde{\Delta}_{c}|^{2}}\cdot\frac{1}{|1-v|^{2}},\label{eq2}
\end{equation}

where
\begin{equation}
v=\frac{2g^{2}}{\tilde{\Delta}_{c}[\tilde{\Delta}_{a}-\tilde{J}]}\label{eq2}
\end{equation}

These results are based on the classical approximation~\cite{Schuster},
which converts the fermionic
commutation relation to a bosonic form, treating  both atoms and field
as linear harmonic oscillators and  omitting the effects of saturation.

The two normal-mode resonances are characterized by the eigen-frequencies $\omega_{\pm}$,
\begin{equation}
\omega_{\pm}=-\frac{1}{2}(\tilde{\Delta}_{a}-\tilde{J}+\tilde{\Delta}_{c})\pm
\frac{1}{2}\sqrt{[8g^{2}+(\tilde{\Delta}_{a}-\tilde{J}-\tilde{\Delta}_{c})^{2}},\label{eq2}
\end{equation}
which are based on the Eq.(9). The frequencies $\omega_{\pm}$ have complex values.
The real part Re($\omega_{\pm}$) determines
the position of  the resonances, while the imaginary part describes their widths.
However, in the strong-coupling regime, $g\gg(\gamma, \gamma^{'}, \kappa)$,
so the effects of the decay on the position of  the resonances can be neglected.
The resonance frequencies $\omega_{\pm}$, in this condition, match the first pair of dressed states in Eq.(11).
For $\Delta=J=0$, the distance beween the two resonances has a minimum value
$\omega_{+}-\omega_{-}\approx2\sqrt{2}g$. When $\Delta$ and $J$ are nonzero,
the position and distance of  them can be obtained from the expression of Eq.(11).

As the two-atom system to some extent  is analogous to the one atom system,
we cite the parameters value in~\cite{Birnbaum}, in which photon blockade for
the light transmitted by an optical cavity containing one trapped atom
is observed. The decay rate $\gamma'$ due to the DDI is usually weak,
so without loss of generality, we take it as $0.05g$ in this section.

\begin{figure}
\begin{center}
\includegraphics[width=0.7\textwidth]{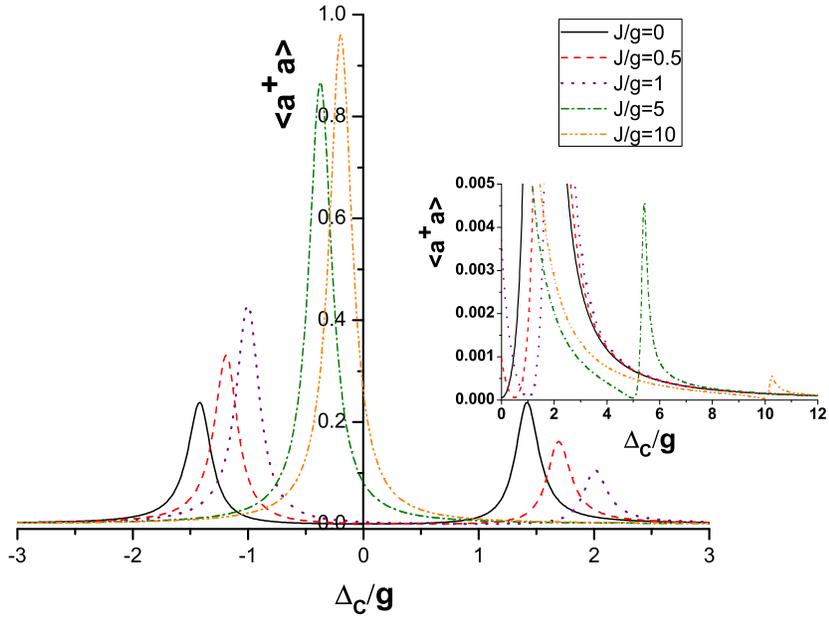}
\caption{\label{switchfunc}
The normal-mode spectrums  for different DDI intensities is shown.
The parameters are $\Delta=0$,
($\eta$, $\kappa$, $\gamma$, $\gamma^{'}$)=(0.12, 0.12, 0.0767, 0.05)$g$.}
\end{center}
\end{figure}

In Fig.1, the normal-mode spectrums  for different DDI intensities is plotted.
From Fig.1 we can find that when $J$$=0$, the amplitudes of both resonances are equal
and the position of  them is symmetric about  $\Delta_{c}$$=$$0$ with a minimum
distance about $2\sqrt{2}g$. However, when   $J$$\neq$$0$, it indicates that with
the increase of DDI intensity, the left peak bacomes higher and gets closer
to $\Delta_{c}$$=$$0$, while the height of right peak is reduced greatly
and it gets far away from $\Delta_{c}$$=$$0$. Moreover, the distance between
the two peaks shows  an  enlarged tendency.
This is because for $\Delta=0$, $J\neq0$ the distance between the two peaks
is in accord with the expression $\sqrt{J^{2}+8g^{2}}$, which is a
monotone increasing function of $J$.
In addition, the height of them are determined by the value of  $\sin\frac{\theta_{n}}{2}$
and $\cos\frac{\theta_{n}}{2}$  in Eq.(10).
When $\Delta=J=0$, the contributions from the atoms and the cavity states
are equal so that the normal modes have the same height.
With the increase of DDI intensity, the excitation probability of
the bare cavity field  state  enhances
for the lower dressed state $|1_{-}\rangle$,
while  for  the bare atoms states it reduces greatly.
The results are opposite for  the higher  state $|1_{+}\rangle$.
It should  be noticed that the system is pumped by a coherent laser  beam shining
on one of  the cavity mirrors, so the bare cavity states are more easily  excited,
leading to a  better visibility of  the  ``cavity-like" peak.
However, when $J\gg g$, the probabilities $\sin^{2}(\frac{\theta_{n}}{2})\approx0$,
$\cos^{2}(\frac{\theta_{n}}{2})\approx1$, so the system almost is in the state $|g,1\rangle$.
The atoms are not being excited in this case  and the spectrum shows a single peak.

\begin{figure}
\begin{center}
\includegraphics[width=0.7\textwidth]{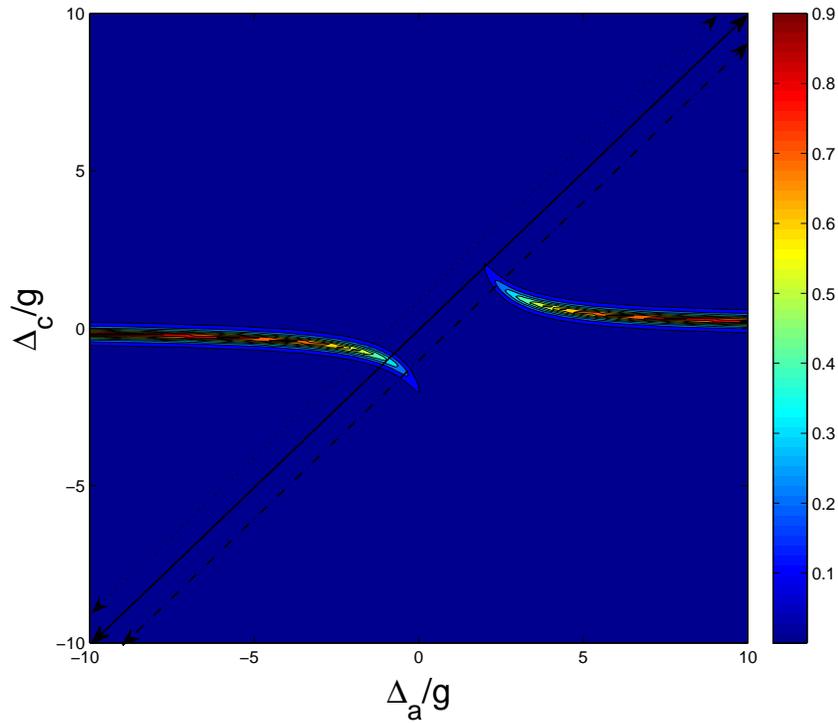}
\caption{\label{switchfunc}
The normal modes form an avoided crossing between the resonances of the bare atoms and the bare cavity.
Three different cases  of $\Delta/g=1$(dashed-dotted), $\Delta=0$(solid) and $\Delta/g=-1$(dotted) are shown.
The dipole-dipole interaction strength is taken as $J/g=1$.
The parameters are
($\eta$, $\kappa$, $\gamma$, $\gamma^{'}$)=(0.12, 0.12, 0.0767, 0.05)$g$.}
\end{center}
\end{figure}

In Fig.2, the normal modes spectrum are revealed as  functions
of $\Delta_{a}$ and $\Delta_{c}$. The atom-cavity detuning is
defined as $\Delta=\omega_{a}-\omega_{c}$.
An avoided crossing between the atomic transition and the resonant frequency
of the cavity is shown.
With the  increases of $\Delta_{a}$ and $\Delta_{c}$,
the atom-cavity system decouples and the two resonances approach
the eigenfrequencies of the atoms  and the cavity asymptotically .

\begin{figure}
\begin{center}
\subfigure[]{ \label{fig:subfig:a}
\includegraphics[width=0.7\textwidth]{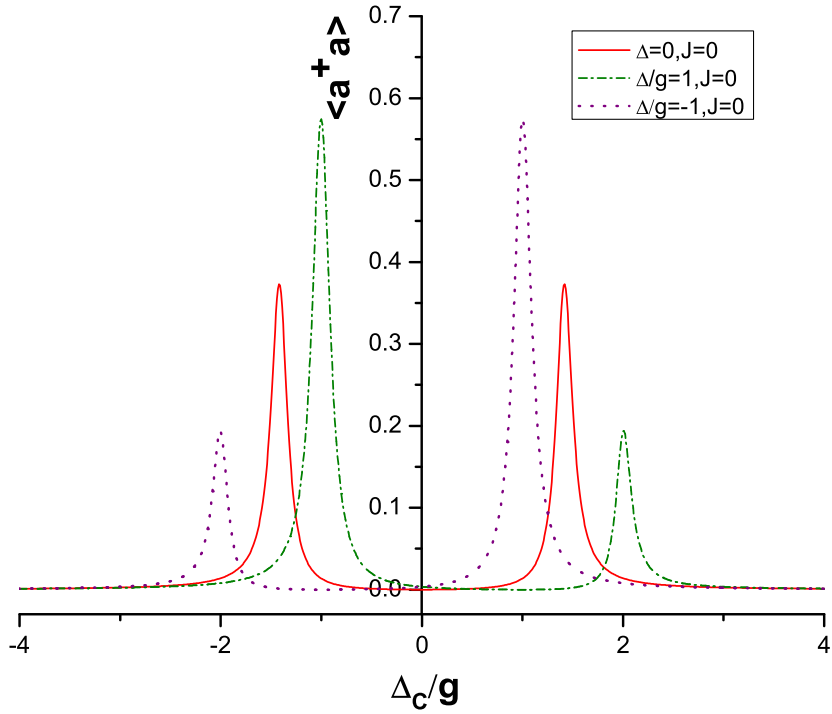}}
\subfigure[]{ \label{fig:subfig:b}
\includegraphics[width=0.7\textwidth]{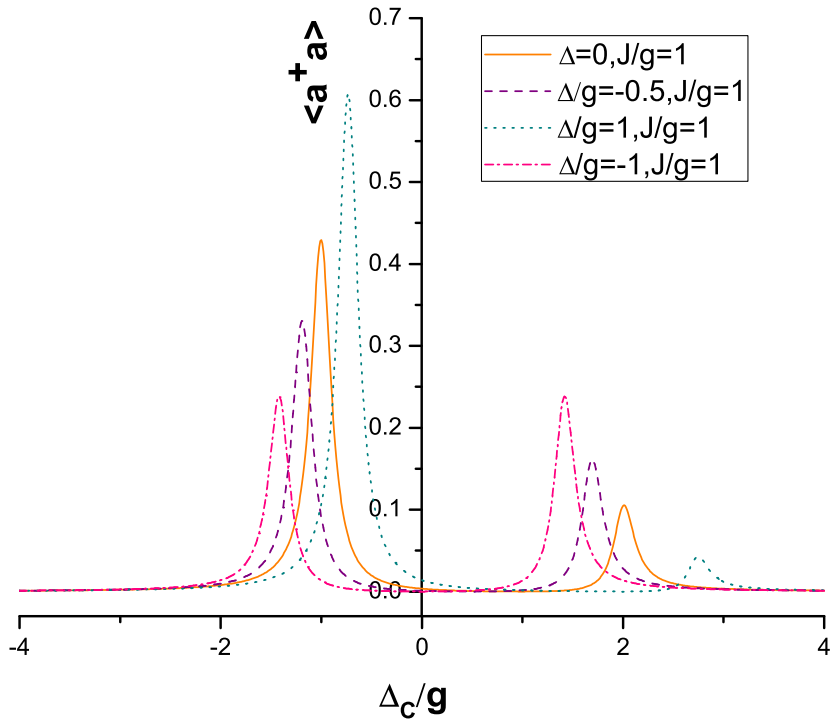}}
\caption{\label{switchfunc}
The normal-mode spectrums for different values of $\Delta$ and $J$ are plotted.
(a)There is no dipole-dipole interaction between atoms, that is $J$=0, $\gamma^{'}$=0.
(b)Both the detuning and the dipole-dipole interaction  are considered.
The dipole-dipole interaction intensity is taken as $J/g=1$.
Other parameters are
($\eta$, $\kappa$, $\gamma$, $\gamma^{'}$)=(0.12, 0.12, 0.0767, 0.05)$g$.}
\end{center}
\end{figure}

In Fig.3(a), when $\Delta\neq0$ both the position and  height of the two resonances shift.
On the one hand, according to Eq.(11), when $J=0$, $\Delta\neq0$, is it obvious that
the position of  the two peaks  rely on the value of $\Delta$ and the  distance
between them depends on  $\sqrt{\Delta^{2}+8g^{2}}$.
On the other hand,  the excitation of a dressed state  is determined by
the contribution of the cavity state to the dressed state.
For positive detuning, based on Eq.(10), the excitation probability of the bare cavity field state
of $|1_{-}\rangle$  augments with the increase of $\Delta$, but for $|1_{+}\rangle$ it reduces.
However, the results are opposite for negative detuning.
Therefore, when $\Delta\neq0$ the two resonances are better to exhibit ``cavity-like" form
with enlarged separation.

Interestingly, as is reflected in Figs.1 and 3(a), it indicates that
the DDI plays a similar  role as the positive detuning.
The effect of the DDI
is equivalent to increasing the  positive detunings and decreasing
the negative detunings. To confirm this conclusion, in  Fig.3(b)
the cooperative action of  $\Delta$ and $J$ on the two resonances is revealed.
It shows that the two resonances are symmetry with equal height
when  $\Delta+J=0$. For $\Delta>0$, $\Delta$ and $J$ have consistent
effects on the two peaks. While for $\Delta<0$, the influences
of them cancel each other, and the practical states  rely
on the larger absolute value one of them.
In fact, these results are apparent.
Both for the dressed states in Eq.(10) and the corresponding eigenenergies in Eq.(11),
$\Delta$ and $J$ are present with the form ``$\Delta+J$".
Therefore,  the position and height of  the two peaks,  as well as the distance between them
all depend on the cooperative action of $\Delta+J$.

\section{Transmission spectrum for higher pump intensity}

We have remarked that the validity of above results depends
on the assumption of the weak excitation.
For higher pump intensity, the atomic saturation can not be neglected.
We can define $\langle\sigma_{z}\rangle_{0}=-\frac{1}{1+s_{0}}$,
and treat the cavity field classically by
replacing $a$ with $\langle a\rangle$.
Then   $\langle a\sigma_{z}\rangle$ can be written as a product form
$\langle a\rangle\langle\sigma_{z}\rangle$
and the steady state of Eqs.(3),(4) and (5) can be calculated.
\begin{equation}
\langle a\rangle_{0}=\frac{\eta}{\tilde{\Delta}_{c}}\cdot\frac{1}{1-\mu}\label{eq2}
\end{equation}

\begin{equation}
\langle\sigma\rangle_{0}=\frac{\eta v}{g}\cdot\frac{1}{1-\mu}\label{eq2}
\end{equation}

\begin{equation}
\mu=\frac{2g^{2}}{\tilde{\Delta}_{c}[\tilde{\Delta}_{a}(1+s_{0})-\tilde{J}]}\label{eq2}
\end{equation}

\begin{eqnarray}
s_{0}&=&\frac{2g^{2}(1+s_{0})^{2}\langle a^{\dagger}a\rangle_{0}}{|\tilde{\Delta}_{a}(1+s_{0})-\tilde{J}|^{2}}\nonumber\\
&+&\frac{2g^{2}(1+s_{0})\langle a^{\dagger}a\rangle_{0}\gamma^{'}}{|\tilde{\Delta}_{a}(1+s_{0})-\tilde{J}|^{2}\gamma}\label{eq2}
\end{eqnarray}
where $s_{0}$ is the saturation parameter. Notice that $s_{0}$$\rightarrow$$0$
corresponds to the low saturation limit.

\begin{figure}
\begin{center}
\includegraphics[width=0.7\textwidth]{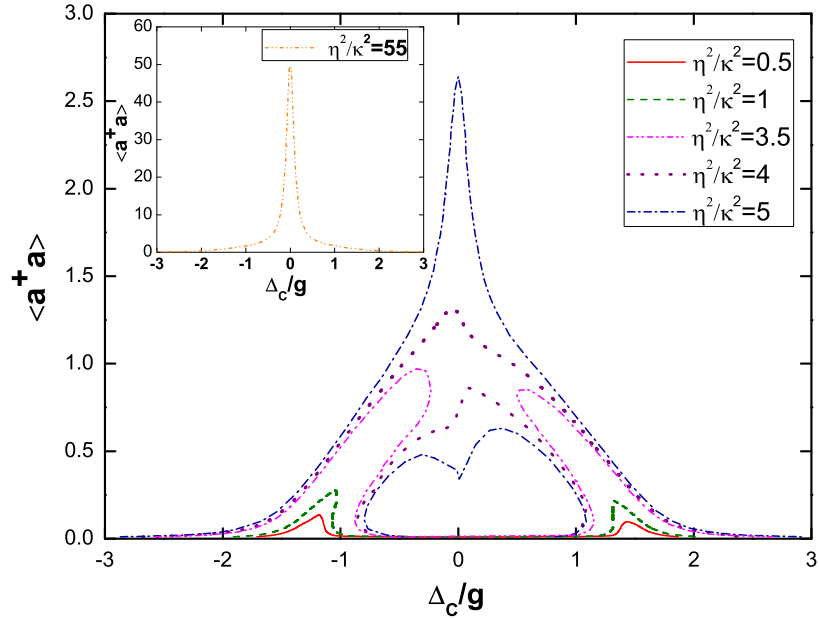}
\caption{\label{switchfunc}
The normal mode structure  is revealed for different  pump intensities.
With increase of pump intensities, atoms tend to saturation,
and  the two peaks bend towards the center.
In the limit of the high excitation, the atoms are saturated and the spectrum
shows a single peak, as is shown in the inset.
$J$$=$$0.5g$, $\kappa$$=$$\gamma$$=$$0.1g$, $\Delta=0$, $\gamma^{'}$$=$$0.01g$.}
\end{center}
\end{figure}

In Fig.4, with Eqs.(19),(21) and (22) the   spectrum
for DDI atoms with increased  pump intensity are plotted.
It shows  similar behavior as the  system without DDI~\cite{Mielke}.
These results are not surprising, because the dipole-coupled
two atoms can be treated as an  effective atom with  renormalizd
atomic frequency and atom-cavity coupling intensity.
With the increase of the pump intensity, atoms begin to saturate
and the peaks of the two resonances shift their position and deform,
bending towards the center. Finally, they meet and  form a closed structure.
It is important to note that
there are three possible expectation values for the operator
$\langle a^{\dagger}a\rangle_{0}$ when $\eta^{2}/\kappa^{2}\geq1$.
One of  them is unstable but the other two are stable.
The amplitude of intracavity field can switch between the
two stable values, named bistability, which predicts a
nonlinear relation between input and output intensity.
The system, in this case, evolves  from two coupled harmonic oscillators
to highly deformed anharmonic oscillators.
In the limit of high excitation, as is shown in the inset,
the atoms are saturated and do not contribute significantly to the
dynamics of the system.
The spectrum resembles that of an empty cavity,
evolving from two peaks to a singlet.
However, the  spectrum no longer shows a symmetrical shapes
as system without DDI.
Because we introduce the DDI into  system,
the shape of these curves are asymmetrical about $\Delta_{c}=0$.

\begin{figure}
\begin{center}
\includegraphics[width=0.7\textwidth]{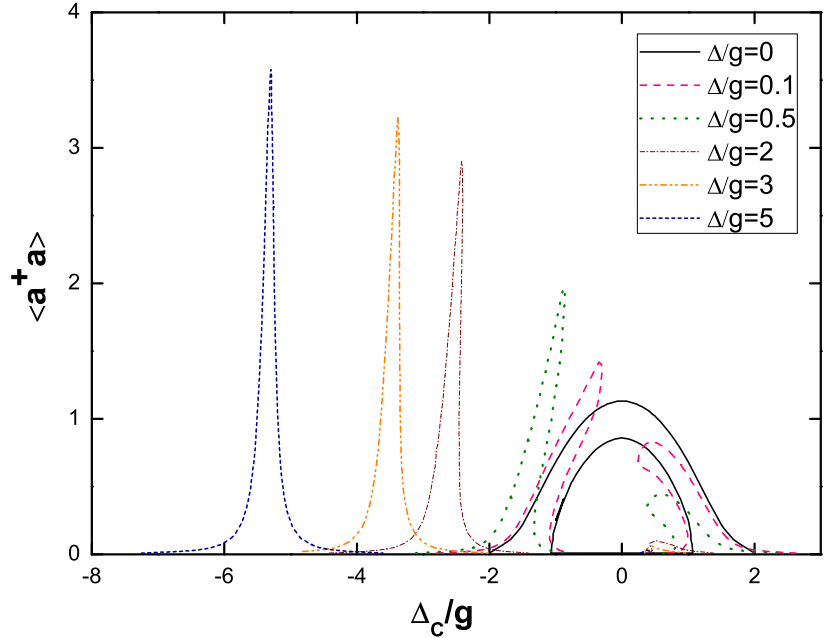}
\caption{\label{switchfunc}
The effects of the atom-cavity detuning on the normal mode structure
of high excitation. The spectrum deformed, and changes to a singlet
in  the limit of large detuning. $\eta^{2}/\kappa^{2}=4$, $J=0$,
$\kappa$$=$$\gamma$$=$$0.1g$, $\gamma^{'}$$=$$0.1g$.}
\end{center}
\end{figure}

In weak excitation limit, the DDI influences both the height and position of the two peaks
and has similar effects as the positive detunings.
So for higher excitation, the effects of $\Delta$ and $J$ on the spectrum are also worth  exploring.
In the following section we begin from the spectrum with closed structure in higher excitation
to study their effects.

In Fig.5, a interesting phenomenon arises
with the increase of  atom-cavity detuning.
The original closed spectrum begins to separate and splits into two peaks.
This can be inferred from Eqs.(19) and (22).
Based on these two  equations, we can get
\begin{eqnarray}
s_{0}=\frac{2g^{2}\eta^{2}(1+s_{0})^{2}(1+\frac{\gamma^{'}}{\gamma})}{|\tilde{\Delta}_{c}[(\Delta_{c}-\Delta+i\gamma)(1+s_{0})-\tilde{J}]-2g^{2}|^{2}}.
\end{eqnarray}
When  $J=0$, with increase of $\Delta$, the atomic  saturation  parameter $s_{0}$ reduces.
It is not difficult to understand that the detuning can make the atom-cavity coupling
becomes weaken, thus  reduces the atomic  saturation.
So with the increase of $\Delta$,
the spectrum is gradually returning  to the cases of weaker excitation.
When $\Delta/g\simeq5$ the right peak nearly vanishes, while
the right peak is more distinct. The system, in this case,
is mainly dominated by $\Delta$ and shows a  decoupled tendency.

\begin{figure}
\begin{center}
\includegraphics[width=0.7\textwidth]{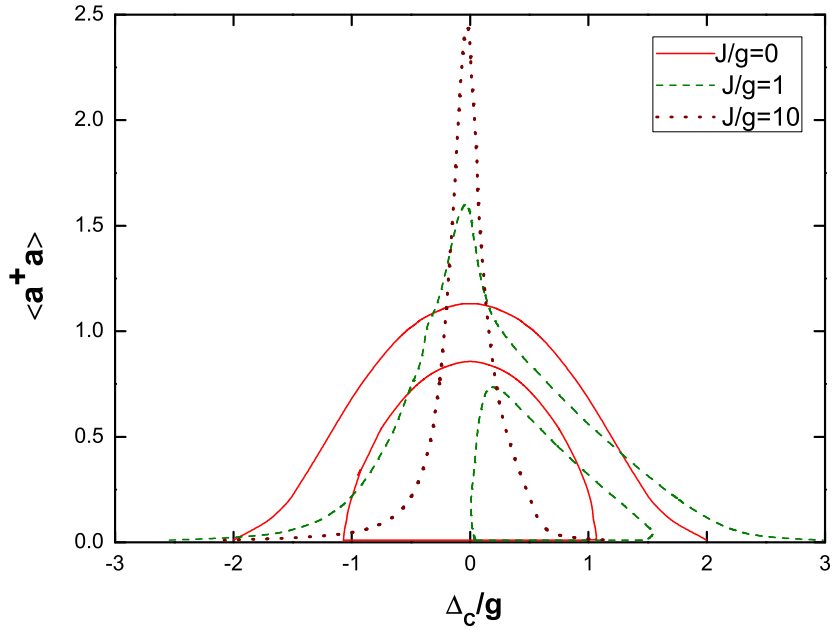}
\caption{\label{switchfunc}
The effects of dipole-dipole interaction on the normal mode structure
of high excitation. The spectrum deformed, and changes to a singlet
in  the limit of high dipole-dipole interaction intensity.
$\eta^{2}/\kappa^{2}=4$, $\Delta=0$,
$\kappa$$=$$\gamma$$=$$0.1g$, $\gamma^{'}$$=$$0.1g$.}
\end{center}
\end{figure}

However,  in Fig.6 the spectrum   as a function of $J$  for higher excitation
presents a different behavior.
The original closed spectrum only deforms but does not separates.
This result maybe explained as follows.
As mentioned above, the DDI can make the atomic frequency renormalized
and change the atom-cavity coupling intensity from $g$ to $\sqrt{2}g$.
According to Eq.(23),  $s_{0}$ increases with the increase of $J$
for $\Delta=0$. For high excitation, the renormalization of the atomic
frequency resulting from the small DDI does not generate pronounced
influences on the excitation of the system, then the spectrum only deforms.
However, for $J\gg g$ cases,  a large atom-cavity detuning forms due to
the renormalization of the atomic frequency and plays a dominant role
in the atom-cavity system. Then photons coming from a cavity mirror
can not populate on the atomic states, making the spectrum shows a singlet.

\section{Conclusion}

We have characterized the transmission spectrum properties of two dipole-coupled two-level
atoms  strongly coupling to a single-mode optical cavity.
In the low excitation limit, the DDI, acting as the atom-cavity detuning, can change
the position and height of the two peaks. However, the DDI has similar
effects as the positive detuning, and shows opposite effects as the negative
detuning. The dressed states  have also been   derived by transforming the
two-atom system to an effective single-atom system.
For higher excitation,  the atom-cavity detuning  can  reduce the atomic
saturation, making the original closed structure separate.
While the DDI can augment the atomic saturation,
leading to the deforming of the original closed structure.
Interestingly, except for the limit of high excitation and
large detuning, the  strong DDI can also result in the decoupling
of the atoms and cavity, which  leads to the spectrum showing  a singlet.
We expect that these results will be useful in  understanding
the quantum electrodynamics of the atom-cavity system with DDI.

\section{Acknowledgement}

This work was supported by NSFC under grants
Nos. 10704031, 10874235, 10934010 and 60978019,
the NKBRSFC under grants Nos. 2009CB930701,
2010CB922904 and 2011CB921500, and FRFCU under
grant No. lzujbky-2010-75.

\end{document}